\documentclass[11pt]{amsart}
\usepackage[foot]{amsaddr}
\usepackage{graphicx}
\usepackage{epstopdf}
\usepackage[sort&compress,numbers]{natbib}

\begin{document}

\title{Dynamics and Processing in Finite Self-Similar Networks}
\author{Simon DeDeo$^\dagger$}
\author{David C. Krakauer$^{\ddagger,\star,\dagger}$}
\address{$\dagger$ Santa Fe Institute, Santa Fe, NM 87501, USA}
\address{$\ddagger$ Wisconsin Institute for Discovery, University of Wisconsin, Madison, WI 53706, USA}
\address{$\star$ Department of Genetics, University of Wisconsin, Madison, WI 53706, USA}
\email{simon@santafe.edu, krakauer@santafe.edu}
\date{\today}

\begin{abstract}
A common feature of biological networks is the geometric property of self-similarity. Molecular regulatory networks through to circulatory systems, nervous systems, social systems and ecological trophic networks, show self-similar connectivity at multiple scales. We analyze the relationship between topology and signaling in contrasting classes of such topologies. We find that networks differ in their ability to contain or propagate signals between arbitrary nodes in a network depending on whether they possess branching or loop-like features. Networks also differ in how they respond to noise, such that one allows for greater integration at high noise, and this performance is reversed at low noise. Surprisingly, small-world topologies, with diameters logarithmic in system size, have slower dynamical timescales, and may be less integrated (more modular) than networks with longer path lengths. All of these phenomena are essentially mesoscopic, vanishing in the infinite limit but producing strong effects at sizes and timescales relevant to biology.
\end{abstract}

\maketitle

Biological networks exhibit a wide range of structural features at multiple spatial scales \cite{Strogatz:2001p16837,Newman:2006p16621,Babu:2004p16485}.  These include local circuitry reflecting the logic of regulation among small numbers of elements \cite{ED2006}, and motifs of statistically over-represented patterns within larger networks of interactions \cite{Alon:2007p16472}, through to macroscopic properties of complete networks including the description of the degree distributions and the large scale geometric features of networks~\cite{Newman:2006p16621}. Among the most interesting geometric properties of biological networks is the property of self-similarity or scale invariance~\cite{Song:2005p13358,Albert:2005p16492}, in which characteristic topological features are present at all scales from the local organization of individual nodes, through to aggregations at the largest network scales. 

For genetic and proteomic regulatory networks, as well as social networks and a variety of distribution networks, including respiratory and circulatory networks, the mechanisms generating self-similar structures have been well explored~\cite{Goldberger:1992p16245,Orbach:1986p16265,gr1,gr2,Ravasz30082002,human,human2,Hamilton07092007,BRV:BRV192,jen}. A growing body of empirical work investigates self-similar network structures, including motif overabundances at different coarse-grained scales. The topology of networks under coarse-graining (agglomeration) of nodes has formed a central focus in both empirical~\cite{Song:2005p13358} and theoretical~\cite{Rozenfeld:2007p20583,Rozenfeld:2007p20584,Radicchi:2008p13331,Radicchi:2009p17143,Rozenfeld:2010p20551,Bizhani:2011p20547,Bizhani:2011p20548} work. However, the functional implications of these topological properties remain poorly understood.

Functional explanations of self-similarity tend to fall into one of three broad classes. Robustness explanations consider the connectivity properties under perturbation, and contrast, for example, scale-free and exponential degree distributions~\cite{Estrada:2006p16608,Dekker:2004p16619,Callaway:2000p16611}. Adaptive optimization theories argue that self-similarity provides an efficient means of provisioning densely distributed resource sinks with a minimum of cable cost \cite{West:1999p16260,Gallos:2007p16251}. Hence networks such as the circulatory system can efficiently provide energy-rich compounds to the cells of the body, and neural networks can efficiently integrate information from a large variety of sensory inputs \cite{Stevens:2001p16716,Chklovskii:2002p16715}. 

Finally, neutral theories suggest that self-similarity is not in itself an optimized property of biological networks, but a consequence of highly conserved developmental processes with local rules of assembly that generate characteristic macroscopic properties \cite{Raval:2003p16927,Evlampiev:2006p16894,Barabasi:1999p16796,Whitesides:2002p13047}. Mathematical studies have shown how motif abundances can be the consequences of constraints on large scale topological properties~\cite{Vazquez:2004p19571}; conversely, large-scale topological features might arise from constraints on a single local property~\cite{Park:2004p8711}. In either case, the connection between the large and small-scale properties of a network may have emerged first without functional meaning.

In this contribution we investigate the \emph{functional} implications of self-similar assembly, as the nodes of a  system adjust their internal states in response to their neighbours and in the presence of environmental noise. We find a tension between the small-world properties of a network and the rapidity of the transition to an ordered phase. For a fixed number of vertices and links, self-similar networks with small-world properties tend to show more gradual transitions, both dynamically, and as a function of noise. The nested-hub structure of such networks provides a bottleneck restricting the possible paths to distant parts of the network. By contrast, hierarchical assemblies, characterized by nesting and a more open structure, have a sharper transition to the ordered state.

Our most surprising results show that while there is some advantage to small-diameter, small-world networks in the high-noise regime, a completely different architecture -- that of nested networks, which eliminates bottlenecks at the expense of longer average paths -- provides greater integration in the low-noise regime. Further, small-world networks produced by branching show dramatically longer dynamical timescales than their nested counterparts. Both features of the nested architecture are driven by the presence of multiple paths between points, as we establish both by simulation and by analytic calculation of the graphical structures that underlie the problem.

A central theme of our investigation are the differences between these constructions in the mesoscopic regime: $N\gg1$, but finite. As we shall see, various properties that vanish in the infinite-size limit lead to pronounced differences in behavior at the finite scales relevant to biology. Our use of both analytic and numerical techniques allows us to investigate two distinct regimes relevant to this mesoscopic phase: analytic results describe the finite-size--infinite-time equilibrium, while numerical simulations show the finite-size--finite-time properties, relevant in the case of strong non-equilibrium effects.

\section{Constructing self-similar networks}
\label{constructing}

\begin{figure}
\includegraphics[width=3.275in]{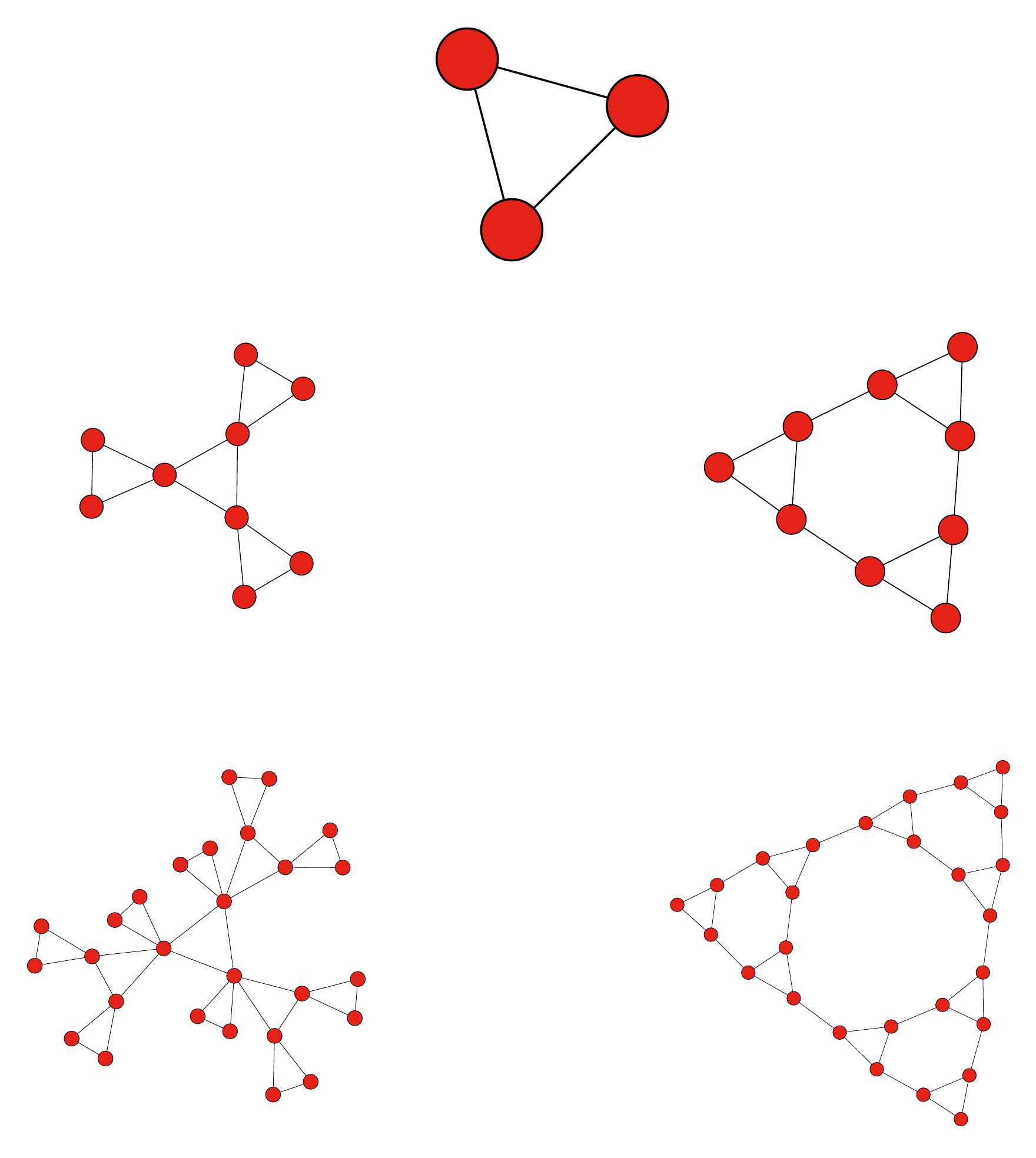}
\caption{Branching (left) and nested (right) iterations on a simple graph.}
\label{triangle}
\end{figure}
We first introduce a deterministic, algorithmic approach for constructing hierarchical, self-similar networks. Our methods use the notion of a construction template, or base motif, that provides the seed for self-similar construction. Alternative stochastic approaches include defining hierarchical assemblies in terms of correlations in an otherwise random network~\cite{Song:2006p13361}, through biases introduced into an ensemble~\cite{Park:2004p8712}, or through high-dimensional generalizations of deterministic constructions~\cite{Zhang:2006p20460,Zhang:2008p20459}. The pseudofractal~\cite{dgm} and the ``flower'' graphs of~\cite{Rozenfeld:2007p20584} are an alternative deterministic construction. An advantage of the deterministic assemblies is that exact calculations of critical behavior become possible. 

The self-similar networks we describe take two forms, depending on their assembly mechanisms -- see Fig.~\ref{triangle}. The assembly mechanism is stated formally in Appendix A; it relies on the specification of (1) a motif pattern $M$ (in Fig.~\ref{triangle}, for example, the triangle), and (2) a method $f$, of replacing nodes in the pattern by new, ``smaller scale,'' copies of the original motif. This method can then be iterated, deterministically, to produce networks of increasing size and complexity.

Visually, our constructions possess fractal-like properties, with self-similarity upon coarse-graining. Our formal definition of the construction of these networks amounts, in the reverse direction, to a specification of a renormalization group transformation~\cite{Nelson:1975p13341}. 

The two simplest choices of node replacement lead to two different kinds of network: a branching topology, characterized by the absence of large-scale loops, and a nested topology, where the loop structure of the base motif is replicated on all scales. We consider the scaling of average network diameter, $\langle d\rangle$, the geodesic distance averaged over all distinct pairs of nodes.

\subsection{Branching Assembly}

As a network grows, a particular unit may preserve the same ``local-structural'' relationship at each level of the hierarchy. For example, the central node of a star may be the central node of the network at all levels of iteration. These networks are characteristic of circulatory and vascular networks, where each node, regardless of its position in a hierarchy, tends to perform the same function~\cite{savage2,Savage28122010}.

In the formalism of Appendix A, such a mapping is provided when $f(i,j)$ is equal to $i$. Iterations increase inequality in the network, producing degree distributions characterized by a motif scale, with a exponential tail of nodes with a ``runaway'' influence on the rest of the system. Biological networks with this property include the neural network of \emph{C. elegans}~\cite{Amaral:2000p13465} and the small-world networks of Ref.~\cite{Watts:1998p13362}. Exponential tails to the degree distribution are found also in the original Erd\"os-R\'enyi random graph.

An illustration of a branching iteration on the triangle is shown on the left column of Fig.~\ref{triangle}. As the order increases, loops, loops with free loops, and so forth are produced; the highest vertex degree increases exponentially in the number of vertices -- these are nodes on the largest ``super-loop.'' All loops, or, in general, subgraphs, may be detached by a single cut. 

\subsection{Nested Assembly}

In contrast to branching iterations, a nested iteration is when the unit -- vertex or subgraph -- takes on the characteristics of its neighbors. A node is no longer restricted to a single local structure, but participates in structures at multiple spatial scales. This is common in communication and computational networks, characterized by extensive feedback loops and connectivity to topologically distinct regions. 

In the formalism of Appendix A, this second mode of network assembly is provided by $f(i,j)$ equal to $j$. While branching structures look tree-like~\footnote{Formally: they have logarithmic scaling of diameter, and no loops on scales above the motif size.}, nested networks are characterized by the replication of subgraph loop structures on increasingly larger scales.



\begin{figure}
\includegraphics[width=3.275in]{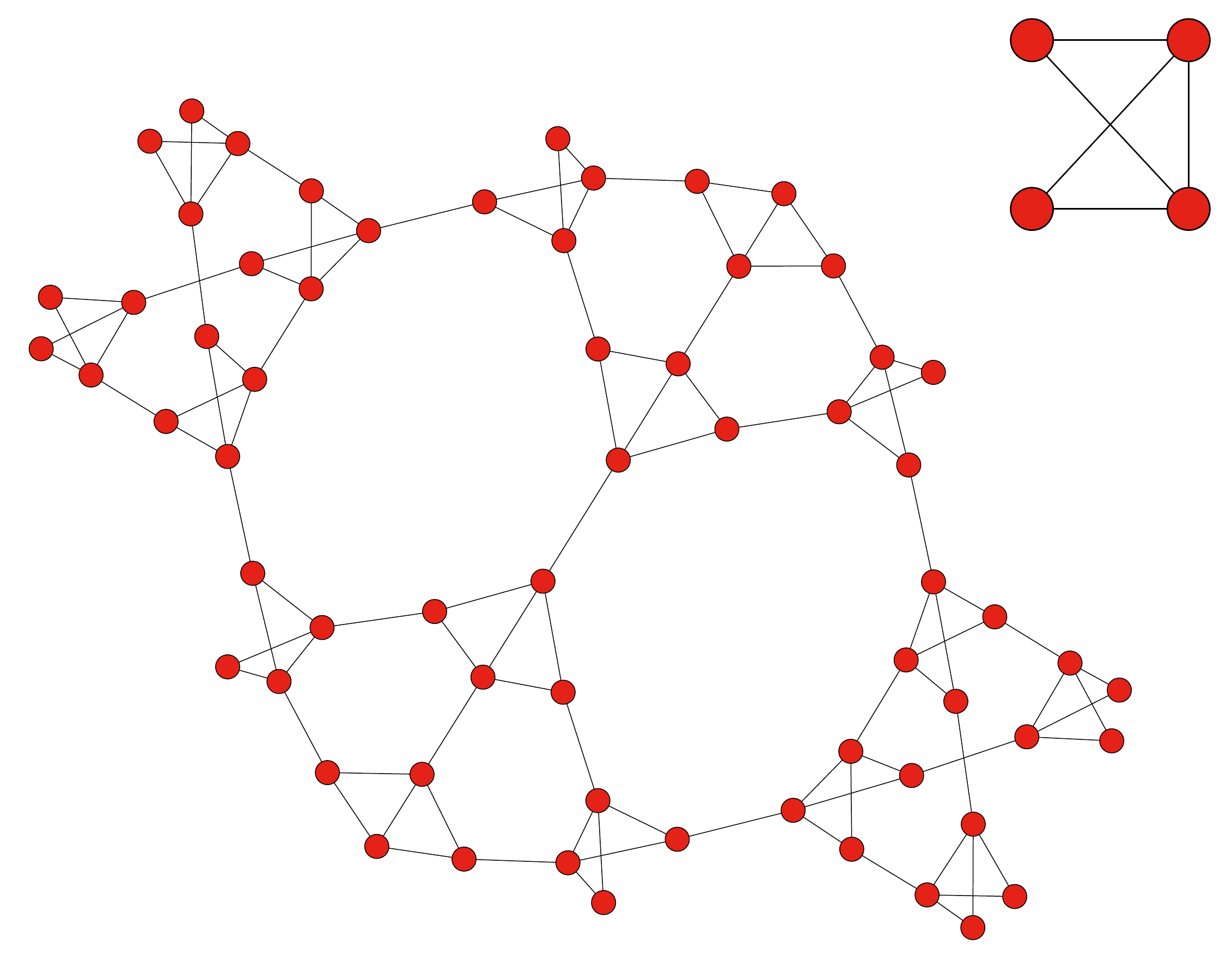}
\caption{Two generations of nested assembly for a common \emph{E. coli} motif; the base graph $M$ is shown in the upper-right corner. The large-scale $\lhd\hspace{-0.78mm}\rhd$ pattern is topologically equivalent to the base motif.}
\label{ecoli}
\end{figure}

\begin{figure}
\includegraphics[width=3.275in]{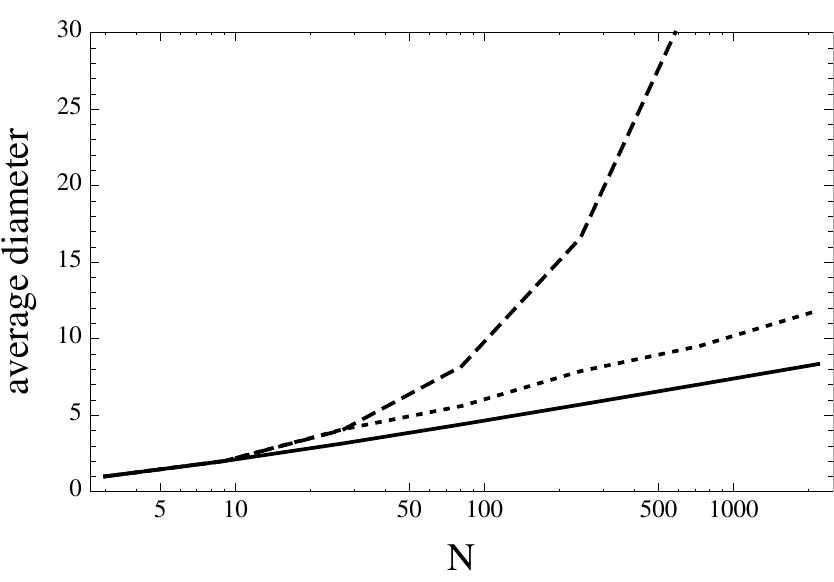}
\caption{``Small world'' behavior in hierarchies. Shown is the scaling of the average diameter, $\langle d\rangle$ with $N$, the number of vertices, for the branching (solid line), nested (dashed line), and mixed (intermediate, dotted line) hierarchies on the triangle motif. Branching hierarchies have the small-world property; $\langle d\rangle\sim\ln{N}$, while nested hierarchies scale as a power law with index roughly that of the motif diameter: $\langle d\rangle\sim N^{\langle d_M \rangle}$. Mixed hierarchies -- shown here, those built of alternating branching and nested iterations -- also scale as a power law, but at a slower rate than the nested case.}
\label{gd}
\end{figure}

Nested networks are shown on the right-hand column of Fig.~\ref{triangle}; a more complicated example is that of Fig.~\ref{ecoli}, where two iterations of a motif overabundant in \emph{E. coli}~\cite{Baskerville:2006p13316} is shown. As shown in Fig.~\ref{gd}, nested graphs have larger diameters; they lack the ``small-world'' property of logarithmic scaling of diameter with size found through replication.

\subsection{Topological Properties}
\label{tp}

The two cases we have considered, pure branching or nesting of a motif pattern $M$, can be considered extremes of how a network might assemble. Branching networks tend to increase inequalities in the degree distributions of vertices while keeping loops at an approximately constant density, whereas nested networks create many more loops while reproducing (almost) the degree distribution of the lower levels.

Another difference between the assembly rules is the scaling of the average diameter. As can be seen in Fig.~\ref{gd}, branching, with its tree-like hierarchy of central hubs, produces small-world graphs where the network diameter scales only as the logarithm of network size~\cite{Watts:1998p13362}. 

This can be understood by considering successive construction steps. The increase in the number of vertices at each iteration leads to an exponential scaling of system size with iteration number:
\begin{equation}
N_i-N_{i-1}=(n-1)N_{i-1},
\end{equation}
where $n$ is the number of vertices in the base motif and $N_i$ is the total number of vertices at iteration $i$. The tree structure of branching networks, however, means that distance between nodes on the perimeter (\emph{i.e.}, those nodes with the greatest separations on the graph) only increase by a constant, proportional to $n$. The diameter, in other words, increases linearly at each iteration, and so the network as a whole has only a logarithmic scaling between diameter and system size.

Diameter increases much more rapidly for the nested structures. Crossing such a structure requires crossing the nested subgraphs, and so the separation between distance points increases proportional to $N_i$ as well as $n$. This leads to a power-law scaling, with diameter increasing as a power of the number of vertices. The index of the average diameter scaling is the average diameter of the base motif. All of these relationships are shown in Fig.~\ref{gd}.

\begin{figure}
\includegraphics[width=3.275in]{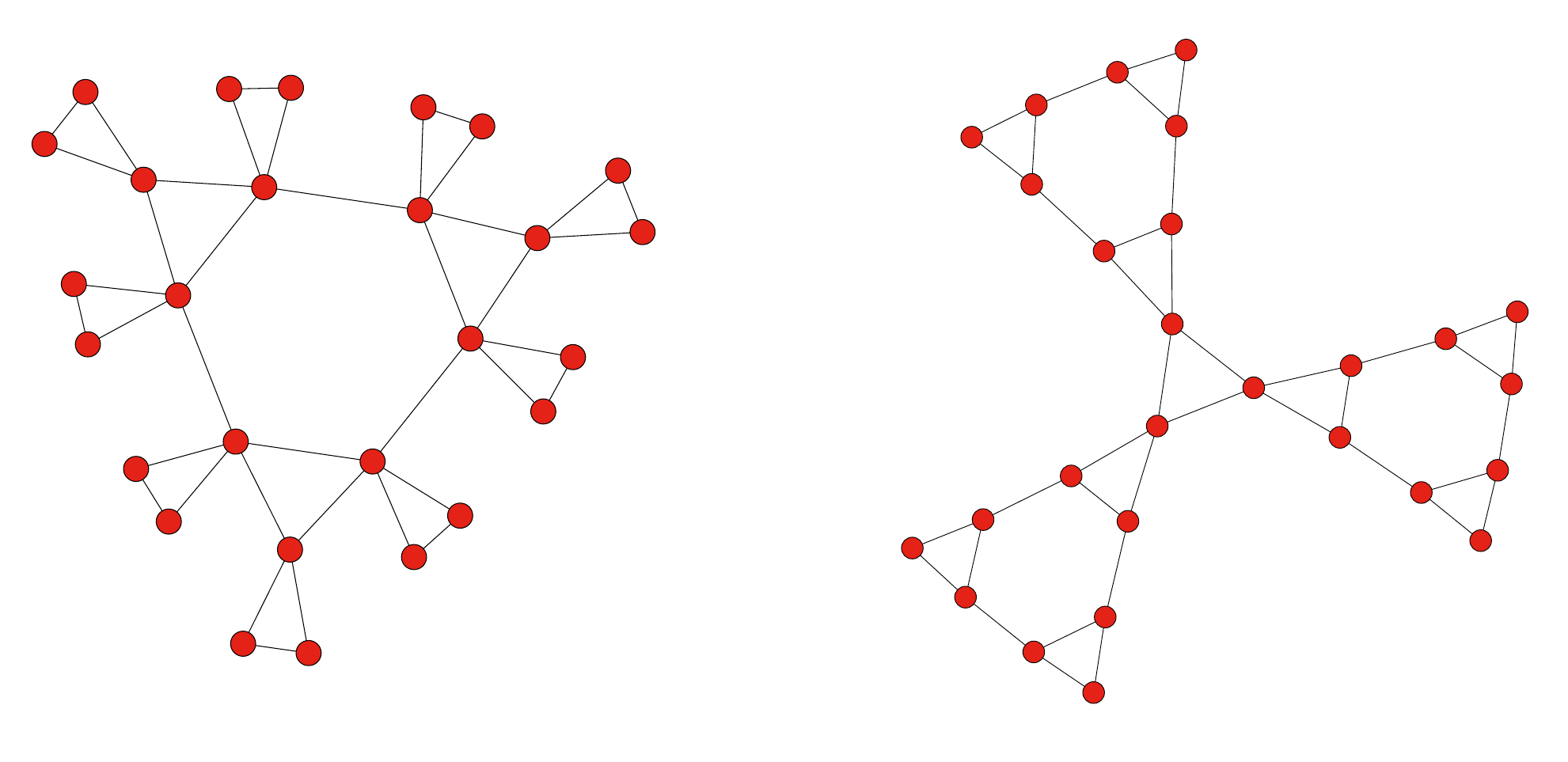}
\caption{Mixed iterations on the three-vertex loop. Branching then nesting (left); nesting then branching (right.) Subsequent iterations define connectivity on increasingly larger scales.}
\label{mixed}
\end{figure}
When two formation patterns are mixed so that a graph may switch from branched to nested, the results is networks as in Fig.~\ref{mixed}. As a means of self-organization, mixed iterations lead to a kind of self-\emph{dis}similarity, where coarse-graining reveals different organizational principles at different levels. Evidence for self-dissimilarity under coarse-graining has been found in both biological and engineered systems~\cite{Itzkovitz:2005p19592}.

\section{Signaling, Modularity and Noise}
\label{signal}

Whereas some network features and motif structures could arise through simple genetic or developmental stochastic processes, non-essential conservation rules, or due to the local constraints of physics and chemistry, we show that two extremes of network structure can still have important functional implications for the ways in which different parts of a network become correlated, or exchange information.

A crucial concept for this work is that of noise, which accounts for the influence of random events and unobserved degrees of freedom in a system. Particular examples of noise might include the small-number fluctuations in reactants that affect metabolic processes, the coupling of observed neurons to part of the larger, unobserved network, or the use of mixed strategies in a game-theoretic system. In the absence of strong theories for the noise properties of a particular case, we use a maximum-entropy model, as described below.

In particular, for our dynamics, we take nodes to have two states (``on'' or ``off'') approximating the discrete switching events observed in a number of systems from the cellular~\cite{Macia:2009p18657} to the social~\cite{DeDeo:2010p18133}. We follow recent work showing the dominance of pairwise interactions in system behavior~\cite{Schneidman:2006p9857,Bialek:2007p9742,Fitzgerald:2009p12175}, and consider networks with pairwise constraints described by a maximum entropy model. This amounts to requiring the full state of the system -- the switch-state of all $N$ vertices -- be given by the Boltzmann distribution. This is then the Ising model on an arbitrary graph.

We can then write the Boltzmann distribution of spins $P(\{\sigma\})$, as given by the set of pairwise constraints $J_{ij}$, and external fields $h_i$,
\begin{eqnarray}
E(\{\sigma\}) & = & \frac{1}{2}\sum_{ij}J_{ij}\sigma_i\sigma_j  \nonumber \\
P(\{\sigma\}) & = & \frac{1}{Z}\exp{-\beta\left[E(\{\sigma\}) + \sum h_i\sigma_i \right]}
\label{ising}
\end{eqnarray}
The $J_{ij}$ are simply the edges of the different networks we consider. The system is in the maximum entropy state with only one observable -- average total energy, or number of satisfied pairs -- fixed~\cite{Jaynes:1957p4563}. Usually, the $h_i$ are taken to be zero; when they are non-zero it amounts to external constraints acting on single nodes -- such as one might expect in a network partially devoted to sensing external conditions.

The most important parameter for this study is the overall factor of $\beta$. Large values of $\beta$ correspond to the low-noise regime; conversely, as $\beta$ goes to zero, the coupling between nodes is swamped by random fluctuations. We refer to $\beta$ as the inverse noise, and focus on how changing $\beta$ leads to changes in how the network correlates and processes information in both equilibrium and non-equilibrium situations.

Determining the correlational and information-theoretic properties of the networks involves finding the joint probabilities of the states of the network, $P(\{\sigma\})$.  In general, we are interested in quantities such as $P(\sigma_i,\sigma_j)$, the joint probability of two nodes $i$ and $j$ being in the same, or opposite, switching states.

There are many different approaches to finding, or approximating, $P(\{\sigma\})$; they are valid in different regimes. For the construction rules we consider, for motif structures with maximum degree of two (\emph{i.e.}, chains), exact solutions of the Ising model are possible via a renormalization group transformation, and for structures with maximum degree of three, an exact solution for the partition function in zero field is generally possible~\cite{Fisher:1959p13350}. For arbitrary motifs, however, the partition function for the $i$th iteration can no longer be written as the partition function for the $(i-1)$th with a suitable change of coupling, $J\rightarrow J^\prime$.

In this paper, we adapt the ``direct configurational'' method (DCM; see, \emph{e.g.}, Ref.~\cite{Oitmaa:2006p13286}), which allows exact computation in small, finite networks. The self-similar properties of our networks allow these computations to be extended to graphs with many hundreds of nodes.

The computation of $P(\{\sigma\})$ can be done via the normalizing term, or ``partition function,'' $Z$, in the denominator of Eq.~\ref{ising}. Derivatives of $Z$ with respect to $h$ then give moments of $P(\{\sigma\})$. These computations not only provide an exact solution, but decompose graphically into sums of ``paths of influence'' closely related to the Feynman diagrams of condensed matter and particle physics. Appendix B describes these calculations, which provide a rigorous basis for the qualitative discussion of how multiple paths lead to critical phenomena (see, \emph{e.g.}, Ref.~\cite{Stanley:1999p13368}.)

\section{Phase Transitions and the Mesoscopic Regime}

Despite the simplicity of Eq.~\ref{ising}, the model has a rich set of behaviors, including (depending on the graph structure) a critical point, $\beta_c$. The characterization of critical phenomena has been a central theme of the study of complex networks~\cite{Dorogovtsev:2008p9412}. With exact expressions for the correlations in hand (see Appendix B), we can  study the nature of the order-disorder transition on the different hierarchies presented here.

Despite the length of the expansions -- ratios of two power series in $\tanh{\beta}$ to $\mathcal{O}(N)$ -- the general behavior of the correlation functions for different networks is similar, with a monotonic rise from the disordered to the ordered state. The leading-order behavior in the high-noise limit as $\beta\rightarrow 0$ is $(\tanh\beta)^{r_\mathrm{min}}$, where $r_\mathrm{min}$ is the shortest distance between the two vertices under consideration. The failure of this approximation is due to the increasing number of paths of influence available, which can allow longer paths to dominate if they increase in number quickly enough to offset the exponential suppression in signal.

For branching networks, the tree-like structure suggests that a phase transition in the bulk is prevented at non-zero noise by the nucleation of boundary spins as happens in the Cayley tree (as distinguished from the Bethe lattice)~\cite{MullerHartmann:1974p13065}. Since nested graphs can also be detached by a constant number of cuts at any iteration -- even when $N\rightarrow\infty$ -- the critical point in the thermodynamic limit is also expected to be zero~\cite{Gefen:1984p13599}, similar to the kind of ferromagnetic frustration found in random graphs~\cite{Svenson:2001p12699}. 

For these reasons, it is thus useful to define a critical point for a finite system without reference to a thermodynamic limit but through the behavior of various correlations that, though never mathematically singular, do show the existence of a transition between two distinct behaviors.

For the particular example of the Cayley tree, Ref.~\cite{Melin:1996p10286} introduced the notion of a cross-over noise, $\beta_g$. Decreasing noise, which pushed $\beta$ above $\beta_g$, was associated with the emergence of non-Gaussianity, a slowdown of dynamics, and glassy behaviors such as aging; the critical noise parameter $\beta_g$ goes to the infinite-size limit ($1/\beta_c$ goes to zero) very slowly (as $\log{(\log{N})}$), so that the thermodynamic limit is not representative even for very large systems. The slow approach to thermodynamic limits is often found in finite ramification structures such as the Cayley tree~\cite{StosiC:2003p13583}.

We investigate these systems below using analytic tools (Sec.~\ref{stationary_aspects}) and numerical simulation (Sec.~\ref{dynamical_aspects}), on networks of size $N\sim300$. Because of the extremely slow scaling discussed above, there is a significant range of system sizes where equilibrium properties do not vary appreciable amounts. This region, which we refer to as the \emph{mesoscopic} regime, is qualitatively different from the infinite size limit. The networks we study here are in this regime -- but so are much larger networks, and system sizes nine orders of magnitude larger, for example, are expected to have properties that differ by only a factor of two.

\begin{figure}
\includegraphics[width=3.275in]{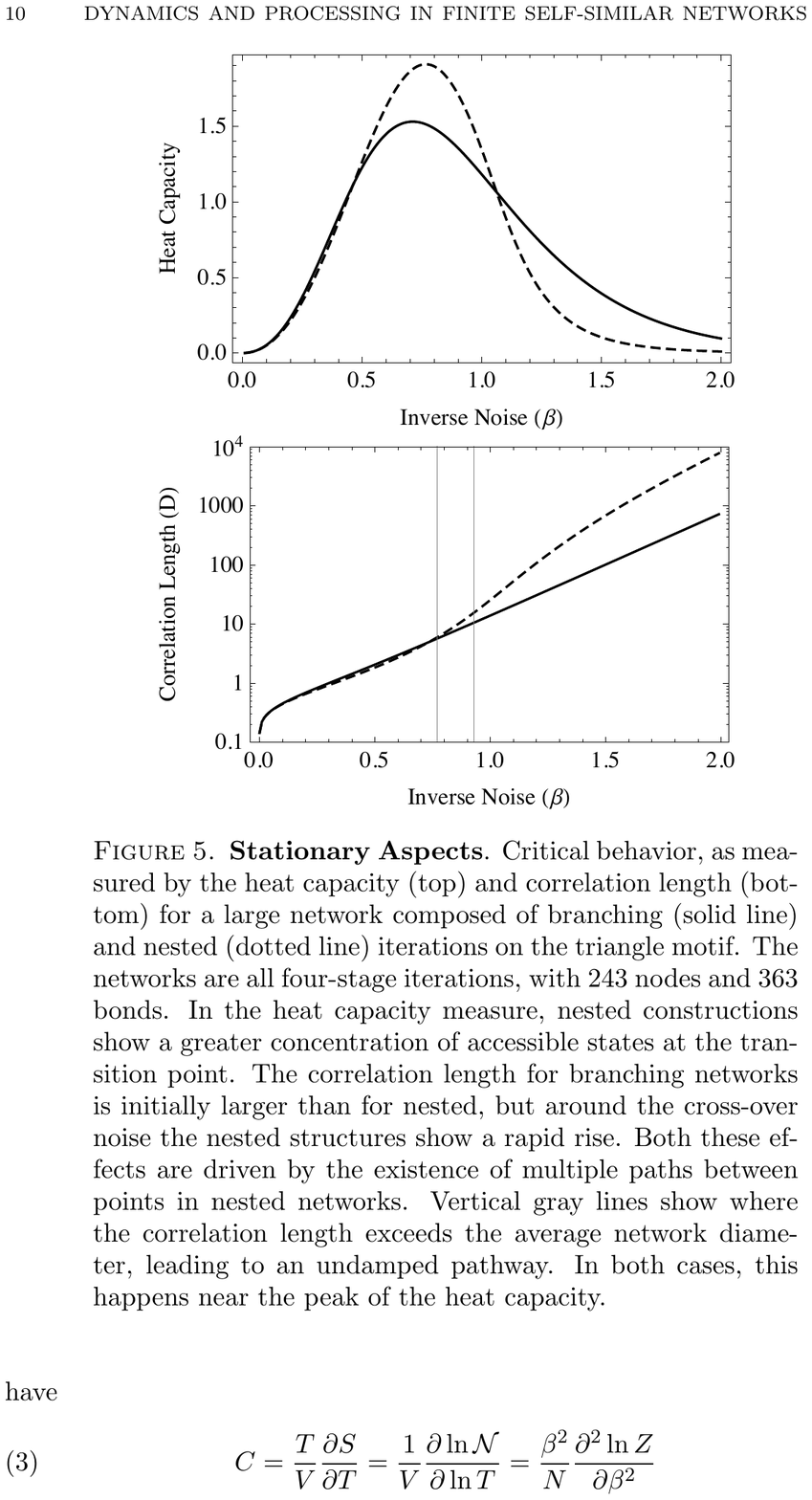}
\caption{{\bf Stationary Aspects}. Critical behavior, as measured by the heat capacity (top) and correlation length (bottom) for a large network composed of branching (solid line) and nested (dotted line) iterations on the triangle motif. The networks are all four-stage iterations, with 243 nodes and 363 bonds. In the heat capacity measure, nested constructions show a greater concentration of accessible states at the transition point. The correlation length for branching networks is initially larger than for nested, but around the cross-over noise the nested structures show a rapid rise. Both these effects are driven by the existence of multiple paths between points in nested networks. Vertical gray lines show where the correlation length exceeds the average network diameter, leading to an undamped pathway. In both cases, this happens near the peak of the heat capacity.}
\label{heat_capacity}
\end{figure}

\subsection{Stationary Aspects}
\label{stationary_aspects}

We measure two quantities related to the stationary, equilibrium properties of the two networks, focusing on their critical phenomena. First we consider the heat capacity, $C$ (see, \emph{e.g.},Ref.~\cite{Tkacik:2006p11059} for an example of its use in biological systems.) At constant external field, we have
\begin{equation}
C = \frac{T}{V}\frac{\partial S}{\partial T}=\frac{1}{V}\frac{\partial\ln{\mathcal{N}}}{\partial\ln{T}}=\frac{\beta^2}{N}\frac{\partial^2 \ln{Z}}{\partial\beta^2}
\end{equation}
where $S$ is the entropy, $V$ the volume (here, the number of nodes), and $\mathcal{N}$ the weighted number of states accessible to the system. The heat capacity measures the (logarithmic) number of states accessible per (log) unit noise. It has a maximum, more or less sharply peaked, as a function of $\beta$. As one heats the system through this point, the number of accessible states increases dramatically, and the intensity and variety of the cooperative phenomena in the transition can be quantified by the height of the peak. Nested networks are characterized by a greater concentration of states, as can be seen in the top panel of Fig.~\ref{heat_capacity}; they can be said to have sharper transitions to the ordered state.

The approach to a phase transition is often defined in terms of a transition between an exponential, and power-law, decline in the correlation function as a function of distance. If we measure the  correlation between pairs of nodes separated by a distance $\Delta r$, we can define a correlation length, $D$,
\begin{equation}
\langle\sigma(r)\sigma(r+\Delta r)\rangle \propto \chi_0^{-\Delta r/D},
\end{equation}
where on these inhomogeneous networks we take $\Delta r$ to be the geodesic distance between points. The bottom panel of Fig.~\ref{heat_capacity} shows the transition that occurs as one passes into the low-noise regime: nested networks, with multiple paths between distant points, allow distant parts of the network to correlate at $\beta\sim1$. 

As noted in Sec.~\ref{constructing}, nested networks have larger diameter. In the case of the $q=4$ iteration, the nested networks have a maximum diameter of 32, compared to 9 for the more tightly structured branching networks. This means that at high noise (small $\beta$), nested networks allow for greater modularity -- distant parts of the network are less correlated. The transition that occurs at $\beta\sim1$, where $D$ for nested networks becomes much larger, reverses this property; nested hierarchies at low noise have stronger long-range correlations. We return to the question of modularity in Sec.~\ref{fluctuation}, where we address it through simulation.

The correlation length $D$ exceeds the average diameter at $\beta$ roughly 0.8 (branching) and 0.9 (nested.) By analogy with infinite-limit, and homogenous, systems, one can consider this noise level as where the effective mass of long-range fluctuations becomes zero. In contrast with the standard infinite limit phase transition, this critical point occurs near, but not precisely at, the maximum of the network heat capacity.

\subsection{Dynamical Aspects}
\label{dynamical_aspects}

\begin{figure}
\includegraphics[width=3.275in]{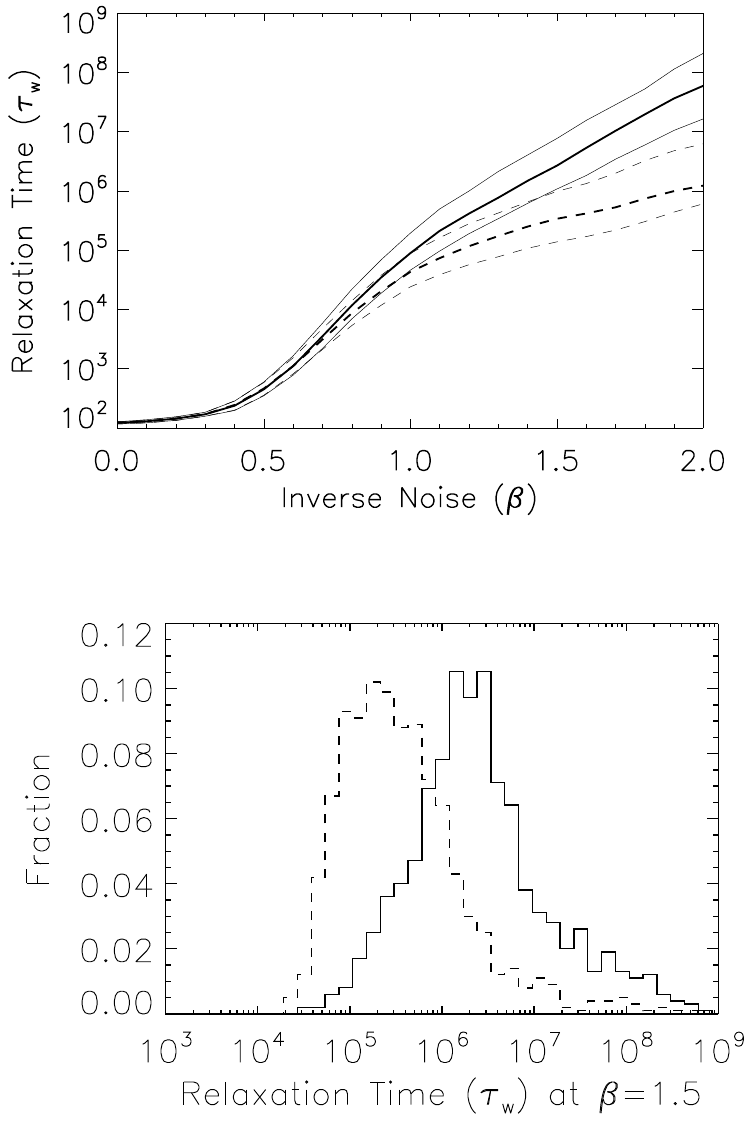}
\caption{{\bf Dynamical Aspects}. \emph{Top}: The relaxation time, $\tau_w$, as a function of inverse noise $\beta$. Timescales are shown for branching (solid line) and nested (dashed line) networks. Network parameters here are the same as in Fig.~\ref{heat_capacity}. As the noise drops ($\beta$ increases), the relaxation times diverge for both networks. At noises below the glassy transition, it is the branching networks that show a stronger slowdown, caused by bottleneck-frustration similar in nature to that of the Cayley tree. Ranges between the thinner lines enclose 50\% of samples. \emph{Bottom}: distribution of $\tau_w$ at $\beta=1.5$, showing the dispersion in relaxation times within a particular network structure.}
\label{overlap}
\end{figure}

Heat capacity and correlation length are both static measures of modularity and signaling. We also expect dynamical signatures of the cross-over in finite networks. In this section, we show that though branching networks are much smaller in diameter (Fig.~\ref{gd}) than nested networks, they have much longer timescales (Fig.~\ref{overlap}.) Nodes are actually less coupled compared to the higher-diameter nested networks, where multiple paths between nodes exist.

In general there are many dynamics compatible with the stationary distributions of Eq.~\ref{ising}. We take the standard Glauber dynamics~\cite{Glauber:1963p12668,sid} with each update step being associated with a different randomly chosen node.~\footnote{We expect other local update rules, such as Metropolis~\cite{Metropolis:1953p17271}, to have similar dynamical properties, with differences appearing only on introduction of non-local rules such as those of Ref.~\cite{nonlocal}.}

Given a sufficiently long time series for any pair of nodes, we can then measure the timescales of their dynamics. We focus here on the decay of the overlap,
\begin{equation}
C(t,t_w)=\frac{1}{N}\sum_{i=1}^N \sigma_i(t)\sigma_i(t+t_w),
\label{overlap-eq}
\end{equation}
where a single step, $\Delta t$ equal to one, is an update of a randomly chosen spin. The function $C(t,t_w)$ decays from unity (at $t_w$ equal to zero) down to (at noises below the glassy phase) a noise floor given by the Poisson statistics of uncorrelated spins. It can be used to measure a number of different properties, including that of aging below the spin glass transition~\cite{Melin:1996p10286}. Here we measure $\tau_w$, the time it takes $C(t,t_w)$ to cross a particular threshold. In Fig.~\ref{overlap}, the threshold is taken to be 0.5, so that $\tau_w$ is the average time for a node to flip with 25\% probability. Because of the long tailed distribution of relaxation times, we follow Ref.~\cite{Billoire:2001p18540} in estimating $\tau_w$ by the median, instead of the mean.

Relaxation times for spin-glass systems are themselves time-dependent -- the longer one lets the system run, the longer the correlation time becomes. This is referred to in the physics literature as `aging'~\cite{dedom} -- correlational properties depend on the age (time since initialization by random initial conditions) of the system. We also see evidence for time-dependent correlation functions past the critical point, similar to that found by Ref.~\cite{Melin:1996p10286}, but focus here on the contrasting behavior of the relaxation time at constant age. We are here in the strongly out-of-equilibrium regime (long timescales on a newly-initialized network.)


The top panel of Fig.~\ref{overlap} shows how $\tau_w$ scales with $\beta$. The strongest differences between the two networks emerge in the low-noise (high-$\beta$) regime. In particular, branching networks, with their hub-and-spoke topologies, show timescales more than two orders of magnitude longer than their nested counterparts.

The differences are due to bottlenecks to communication that exist between distant parts of the network in the branching case. Since all paths between distant nodes must pass through a single point, the speed of communication is limited by the timescale for that single point to change state. 

This is similar to the frustration effects seen in the Cayley tree by Ref.~\cite{Melin:1996p10286}. By contrast, in the nested case there are multiple paths between distant points, and this means that long-timescale frustration effects are limited -- made rarer, though not completely eliminated.

That small world networks, if they rely on hub-and-spoke topologies, are actually slower, is connected to the emergence of long-lived metastable states. The analogs of domain walls for inhomogeneous networks -- separated parts of the system that fall into opposite states of local consensus -- emerge at low noise. These walls propagate through the system until they meet bottlenecks -- places where disparate parts of the network connect via a single node -- and are effectively pinned for long periods. Multiple paths, by contrast, increase the number of points of contact between different neighborhoods.

The dispersion in behavior (lower panel of Fig.~\ref{overlap}) for both networks is large.  This shows the effect of non-equilibrium dynamics and an (effective, since finite-time) breaking of ergodicity. In some cases, the initial conditions, after burn-in, may lead to a particularly ordered configuration from which the system departs with only vanishing probability. In other cases, meta-stable states are not as long-lived, and relaxation can happen quickly.

That one can achieve dispersion in behavioral timescales of nearly five orders of magnitude from a system with only $\sim10^2$ nodes is remarkable. The dispersion, which itself sees an exponential rise at $\beta\sim1$, is another indication of the presence of a finite-size critical phase, present only in the mesoscopic regime.


\subsection{Fluctuation Localization}
\label{fluctuation}

Though branching networks are smaller -- nodes are, on average, closer to each other -- we have shown by simulation that the timescales of dynamical change are much longer (Fig.~\ref{overlap}). Meanwhile, we can determine how many new configurations become accessible as the noise declines from our analytic determination of the heat capacity (Fig.~\ref{heat_capacity}).

In this section, we examine features relevant to information-processing, which is a property of both the stationary properties of the network (how many configurations are accessible) and the dynamical ones (how quickly one configuration turns into another.)

In particular, we ask about the entropy of the system over finite time, and how and where that information is stored: locally (in single nodes), on the motif scale, or non-locally, across widely separated motifs. Such questions are essential to biological function: distinct substructures must not only process information by means of local motif patterns, but also communicate the results of that processing to more distant nodes. Anomalous concentrations of a metabolic product, say, may be detected by influences on one part of the system, but may need to trigger a transcriptional cascade in a different module.

For systems where bits are largely independent, the multi-information is close to zero, indicating that very little information is exchanged between subgroups. When nodes come to process information in complex ways, however, the multi-information becomes larger, indicating that apparent randomness at the local scale becomes pattern exchange on larger scales. Formally, the multi-information at any particular scale is the decrease in entropy seen when the distributions taken by the smaller scales are combined into a joint probability distribution.

We measure the multi-information (see, \emph{e.g.},~\cite{Schneidman:2003p9776}), a generalization of mutual information used to describe cooperative information-processing~\cite{mcgill,Bell:2003p19593}. For the case of three subsystems, whose internal states are represented by a vector, we have for the multi-information
\begin{equation}
I_{nl}=\left(\sum_{i=1}^3 H(P[\vec{x}_i])\right) -H(P[\vec{x}_1,\vec{x}_2,\vec{x}_3])
\label{multiinformation}
\end{equation}

We consider the multi-information between the motif and global scale, with the three the most widely separated, but equidistant triangle motifs chosen as the $x_i$ sets. In words, the first term of Eq.~\ref{multiinformation} is the total entropy of the subsystems considered in isolation of each other; if there were no long-range synchronization, this would be equal to the second term, and the multi-information would be zero. Conversely, since the maximum amount of information in the subsystem is  nine bits, the maximum amount of multi-information is six bits (all three distant motifs perfectly correlated.)

\begin{figure}
\includegraphics[width=3.275in]{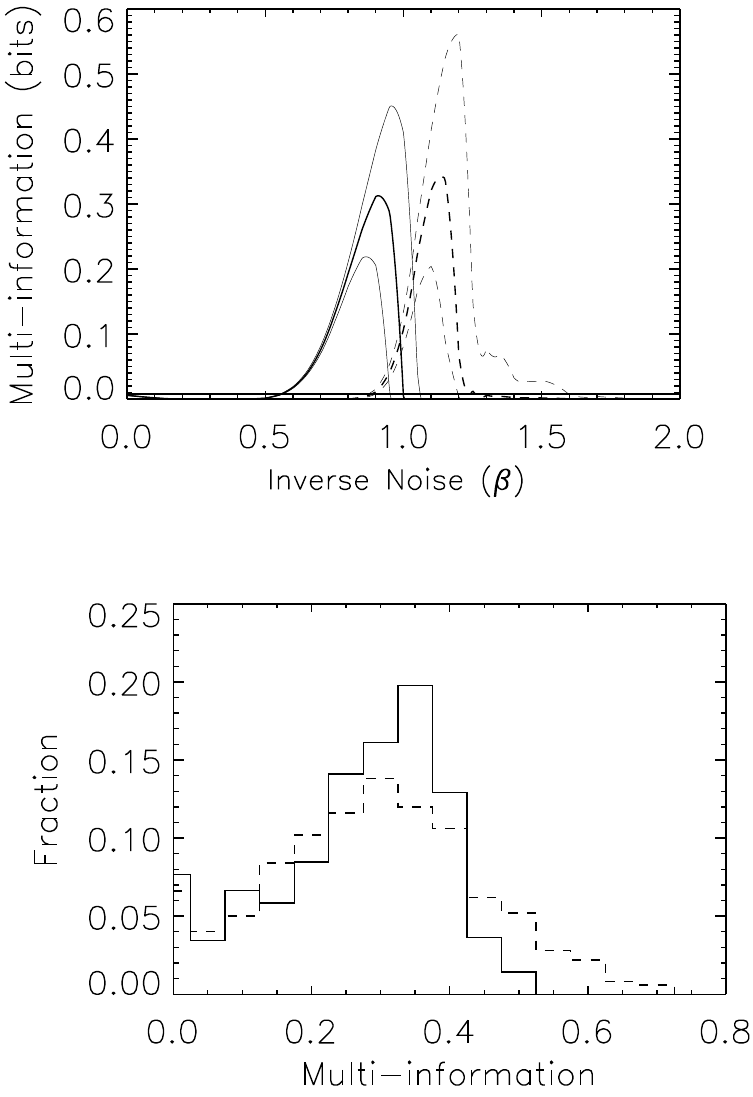}
\caption{{\bf Fluctuation Delocalization}. \emph{Top}: The multi-information between the local (motif-scale) and global scales for branching (solid) and nested (dashed) structures. Also shown are the upper and lower ranges for 50\% of the networks studied. In both cases, greater non-local correlations (high multi-information) are seen as the noise is reduced ($\beta$ increases) -- until a critical point at which multi-information declines to zero, indicating that information processing has become local again. The heavy line at $\approx0.01$ indicates the $1\sigma$ errors associated with the NSB estimator. \emph{Bottom}: distributions near the peak of the multi-information (branching at $\beta=0.9$; nested at $\beta=1.1$) showing the dispersion in measurements.}
\label{graph-multi}
\end{figure}

Fig.~\ref{graph-multi} shows these results, computed directly from simulation. We estimate the multi-information using the NSB estimator~\cite{Nemenman:2001p17550,Nemenman:2004p18597} -- we find it leads to good estimates of simulated datasets with the dramatically lower entropies one expects past the mesoscopic critical points.\footnote{When used to estimate the multi-information by subtraction, we find that the estimator is not unbiased; this effect is overwhelmed, however, for multi-information measurements larger than $10^{-2}$ bits, by the intrinsic dispersion of simulation runs.}

Both the branching and nested structures show a distinct window at which long-range synchronization is strongest and roughly half a bit can be communicated between distant parts of the system. At first, as noise decreases, distant nodes become more correlated (as in Fig.~\ref{heat_capacity}), and the multi-information rises; however, at low noise (large $\beta$), fluctuations on all scales are frozen in. In both cases, this window appears around the same noise-level than the peak of the heat capacity; this provides additional support to the description of a mesoscopic phase transition, since more conventional thermodynamic systems are known to have maximum multi-information at the critical point~\cite{Erb:2004p19378}.



\section{Discussion}

\begin{table}

\begin{tabular}{l|c|c}
& Branching  & Nested  \\ \hline
stationary & & \\ \hline
diameter & small-world & polynomial \\ 
correlations & short distance & long distance \\
phase transitions & soft & hard \\  \hline
dynamical & & \\ \hline
timescales & slow & rapid \\
low-noise processing & local & global \\
\end{tabular}
\vspace{5 mm}
\caption{{\bf Summary of Results}. The behaviors of contrasting self-similar networks in the mesoscopic regime.}\label{summary-table}

\end{table}

In contrast with the regular lattices of field theory, complex networks are characterized by both small-scale pattern and large-scale structural diversity. On small scales, repeating network motifs~\cite{ShenOrr:2002p13402} indicates strong local inhomogeneity. On large scales, networks may be characterized by modularity or by large-scale motifs visible under coarse-graining or aggregation of vertices~\cite{Hartwell:1999p13561,Newman:2006p13774}. The study of such transformations on complex networks has uncovered evidence for self-similarity~\cite{Song:2005p13358}, and small-scale and large-scale network structures, for example, are found to be correlated in cellular networks~\cite{Vazquez:2004p13224}.

In this contribution, we compared two alternative topologies -- networks of branching motifs (with one pattern replicated many times) and networks of nested motifs (where patterns play the role of templates.) Branching networks have a familiar tree-like structure and possess the small-world property; their benefits include efficient signal propagation at high noise. Nested networks retain self-similarity but without small-world scaling, and confer benefits such as redundant paths between distant nodes at the cost of longer path lengths.


A central theme has been the difference at the onset of a mesoscopic version of a phase transition. Phase transitions in general occur in networks when the exponential fading of a correlation along a particular path is balanced by the exponential increase in the number of paths between the two points~\cite{Stanley:1999p13368}. In complex networks, this implies that structural inhomogeneity on a range of different scales will be relevant for the critical behavior analogous to that found in more regular systems.


Our investigation has uncovered a number of counter-intuitive properties of small-world systems. Smaller diameter networks adjust more slowly, have shorter correlation lengths, and can not achieve the levels of non-local integration seen in those nested systems. Our analytic exposition of the problem shows explicitly how the onset of correlations are driven by the existence of multiple paths between points; our simulations show how the existence of such paths allows for the more rapid dissipation of inhomogeneity. Multiple paths are thus central for both information processing and the timescales of coordination. 

In some cases, the characteristic features of the small-world topology listed in Table~\ref{summary-table} are desirable. They can lead to greater modularity, and longer timescales, than they would for more ``open'' topologies with longer path lengths. At low noise, their fluctuations are more localized, meaning that fluctuations in distant structures are increasingly independent, and disjoint memories do not merge and fade as fast. Depending on the nature of computation, these may be desirable properties -- as they are, for example, in the case of the liquid state model~\cite{Maass:2002p19696}.

The existence of such paths also bears on the question of network robustness -- particularly under targeted attack~\cite{Albert:2000p19694}. When all correlational information between two nodes must travel along a single path, the failure of any intermediate node is catastrophic. Conversely, robustness to node deletion will, in general, increase as the number of distinct paths between points increases, even if the number of edges remains constant.

We suggest that our work is particularly relevant to the study of information processing in the brain~\cite{brain,Bullmore:2009p16349}. On the one hand, the maximum entropy model of Sec.~\ref{signal} has formed the basis of a powerful set of models for the description of observed neural correlations~\cite{Stephens:2011p19700}, and the information-theoretic quantities we have investigated are directly related to the Tononi $\phi$ measure~\cite{Tononi:2004p19699} and the $C_N$ measure of Ref.~\cite{Tononi:1994p19594}. These latter measures consider various bi-partitions; the fractal structure of our networks naturally suggest extensions of these measures to the tower of higher-order correlations as described in Ref.~\cite{Schneidman:2003p9776}.

One the other hand, the multi-scale structure of the brain -- from scales of $50~\mu$m to centimeters -- is well-established~\cite[and refs. therein]{Sporns:2006p19595}. The topological and dynamical properties of certain random and deterministic self-similar wirings, relevant to neuroscience, have been under recent investigation~\cite{Sporns:2006p19595,Kaiser:2010p19601}. Our work has direct bearing on explicit models of cortical network architecture~\cite{Robinson:2009p19692}, and in particular suggests that small-world path lengths may not be the only way in which a network might optimize information processing.

Self-similar network properties have proven relevant to the study of a vast range of other natural systems, from gene-regulatory~\cite{gr1,gr2} and metabolic networks~\cite{Ravasz30082002}, all the way up to food webs~\cite{jen} and human social networks~\cite{human,human2,Hamilton07092007,BRV:BRV192}. In the case of social networks, for example, branching networks with complete-graph motifs are small-world examples of the robust social quilts studied by Ref.~\cite{jacksonetal}, while ``span of control'' theories~\cite{md79} address the consequences of hierarchy for information processing and dynamics~\cite{cd09}. Hierarchical structure may also be associated with the emergence of long timescales associated with strategic information processing in animal systems~\cite{DeDeo:2011p18663}.

In parallel, the maximum-entropy models we consider here have proven useful not only in studies of neural functioning, but also in studies of the immune system~\cite{Mora:2010p19701}, and animal behavior~\cite{Bialek:2011p19755,daniels}. In many cases, such systems are found at criticality~\cite{Mora:2011p19702}, making it important to understand the  mesoscopic regime.

The analysis of this paper suggests that statistics related to the existence of multiple paths in a network may be an important way to determine how relevant structural features have been organized to achieve the contrasting properties found in Table~\ref{summary-table}. It may not be necessary to compute all the relevant Feynman diagrams in a graph to answer central questions about the nature of the critical point and ordered phase. When calibrated against the exactly-solved models of this paper, statistics related to the scaling of the number of paths between vertices as a function of distance may be sufficient to study both the nature of the critical point and the existence of non-equilibrium effects. We leave this question for future investigation.

Most theoretical studies have focused on comparing the functional implications of self-similar and non-self-similar networks. We have found none that consider the functional implications of alternative self-similar networks. If it proves to be true that constraints of development account for, and impose, wide spread network self-similarity, then variations on a fractal theme will become the principle means by which development might tinker with functionally important properties. 


\section{Acknowledgements}

SD thanks Van Savage, Tanmoy Bhattacharya, Simeon Hellerman, George Bezerra, and the Institute for the Physics and Mathematics of the Universe, University of Tokyo, Japan, and acknowledges the support of an Omidyar Fellowship. SD and DCK acknowledge support of National Science Foundation Grant 1137929, ``The Small Number Limit of Biological Information Processing.'' DCK acknowledges a John Templeton Foundation award on the Origins and Evolution of Regulation in Biological Systems. 

\newpage

\section{Appendix A: Formal Definitions of Branching and Nested Networks}

Beginning with a motif, $M$, with $N(M)$ vertices, we build up, by iteration, a larger structure, $S(q, M)$, where $q$ is the number of iterations and $S(q=0,M)$ is $M$. The motif directs the assembly of increasingly larger structures, in a recursive fashion, providing the graph with both small and large scale inhomogeneity.

At the $q$th iteration, replace the vertices in $S(q-1,M)$ by separate copies of $M$, and rewire the system while maintaining the local motif structure. One might take the vertices of a triangle, for example, and replace each of them by a copy of the same three-node structure. The different ways to accomplish this model how a network may develop the internal structure of its subsystems; going the other direction, a particular choice defines a coarse-graining operation that might form an element of a renormalization group.

More formally, for each vertex $i$ in $M$, replace the vertex by a copy of $M$, $M_{i}$. For each vertex $j$ in $M$, connect the free edges -- those remaining from the previous iteration that were attached to vertex $i$ -- to the internal vertices of $M_{i}$, by some mapping $f(i,j)$ (generally not symmetric.)

At the $q$th iteration, take $M$ and replace each vertex $i$ in $M$ with copies of $S(q-1,M)$. The rewiring now takes a edge from the $j$th subunit to the $i$th subunit, and attaches it to the $f(i,j)$ vertex in $S(q-1,M)$. The $f(i,j)$ vertex for $S(2,M)$ is defined as the $f(i,j)$th vertex in the $f(i,j)$th subunit, and so forth for higher values of $q$.

Graph $S(q,M)$ has $N(M)^{q+1}$ vertices and $n(M)\sum_{i=0}^q N(M)^i$, or $n(N^{1+q}-1)/(N-1)$, bonds. The average degree of $S$ is always close to that of the local graph $M$, so that sparse networks remain sparse; however, the higher moments of the degree distribution may grow dramatically depending on the choice of $f$.

\begin{figure}
\includegraphics[width=3.275in]{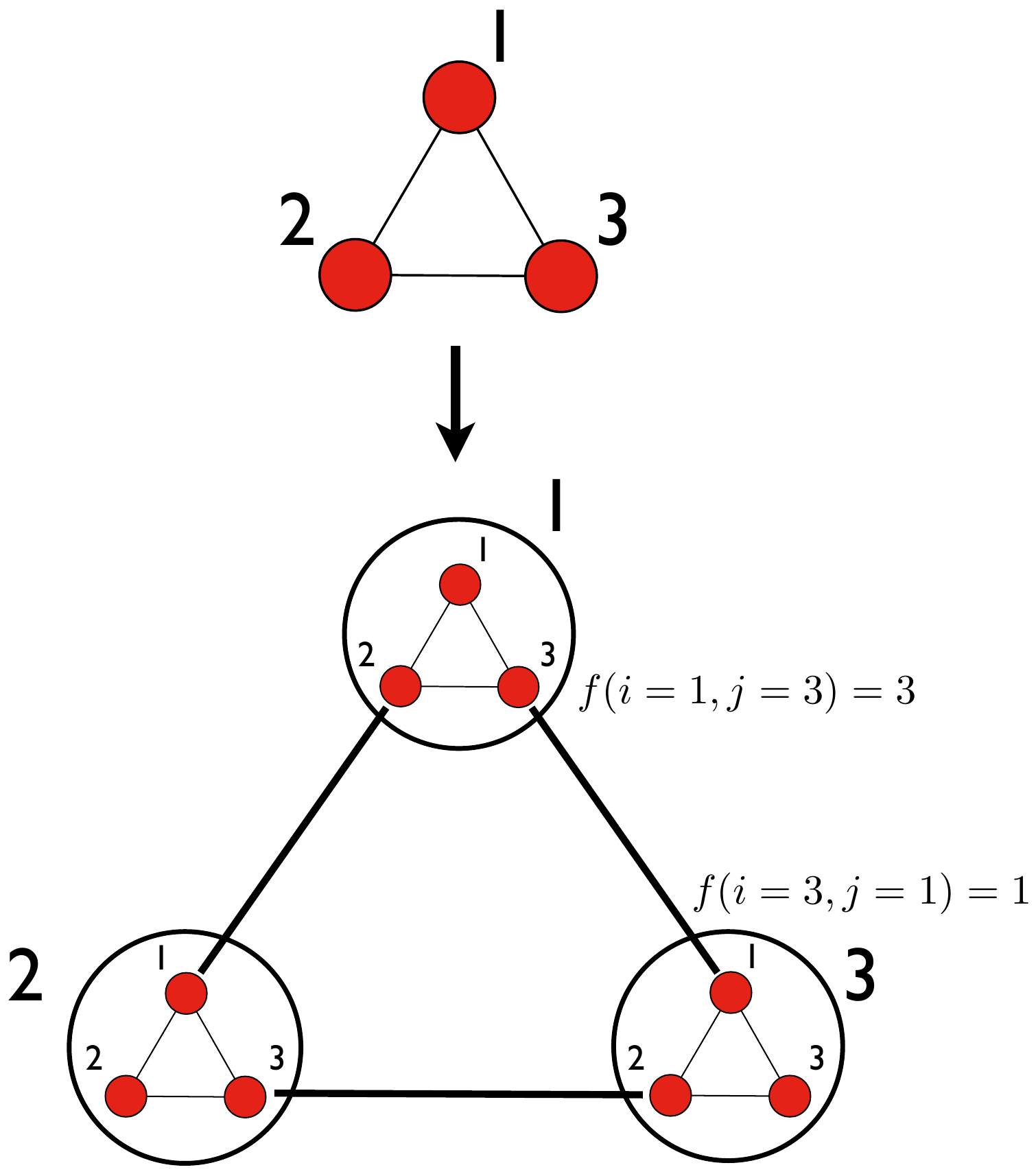}
\caption{An example of how the $f(i,j)$ function specifies which fine-grained node to connect to, for the first iteration of the nested case, $f(i,j)=j$.}
\label{pedagogical}
\end{figure}
Going from $S(q,M)$ to $S(q-1,M)$ is a form of renormalization~\cite{Nelson:1975p13341}. Once $M$ is chosen, the remaining choice is that of the assembly rule, $f(i,j)$; we consider the two simplest cases $f(i,j)$ equal to $i$ (branching assembly), and to $j$ (nested assembly.) These operations are easier to see graphically; for the example of a triangle motif being replicated at multiple scales by the two methods, see Fig.~\ref{triangle}. For a more explicit example of how the $f(i,j)$ rule works, see Fig.~\ref{pedagogical}.

\section{Appendix B: Ising Solutions in the Direct Configurational Method}
\label{a-dcm}

In Sec.~\ref{stationary_aspects} we examined stationary properties of the system. Because of the divergence of timescales discussed in Sec.~\ref{dynamical_aspects}, it is difficult to determine reliable measurements of these properties from simulation. Here we discuss the ``Direct Configurational Method'' (DCM), which allows one to write down expressions for these properties analytically. The expressions are long, but tractable by analytic methods that use computer algebra. They enable us to separate finite-size--finite-time effects (accessible by simulation) from finite-size--infinite-time effects associated with equilibrium.

A number of different expansions for the correlations can be written in the high-noise ($\emph{i.e.}$, $\beta\ll \beta_c$) limit. The most common, known as the linked-cluster expansion~\cite{Wortis:1974p13094}, has formed the center of studies of the Ising model on regular lattices~\cite{MeyerOrtmanns:1997p13040,Reisz:1995p13051,Nickel:1990p12781}. 

Because of the attention paid to lattices with great amounts of symmetry and of infinite extent, less often used are the exact solutions, expressible as a power series with a finite number of terms $(\tanh{\beta J})^n$, available for lattices of finite size. This ``direct configurational method'' (Ch. 2, Ref.~\cite{Oitmaa:2006p13286}) allows one to write an expression for the partition function of a graph directly, by enumerating all of the subgraphs (including disconnected subgraphs) of the original lattice with all vertices even. Similar expressions, with some vertices ``rooted'' in various ways, allow one to determine correlation functions through partial derivatives of $Z$.

For a system of any appreciable size, enumerating the disconnected graphs is a nearly impossible computational task. Finding the free energy, $F$, equal to $\ln{Z}$, turns such a sum of disconnected graphs into a far shorter sum involving only connected graphs, with multiple bonds between vertices allowed, weighted in a new fashion (Ch. 20, Ref.~\cite{Glimm:1987p13287}). These are the usual Feynman diagrams, and allow one to handle an arbitrarily large lattice to finite order in $\beta$. When the lattice has translation symmetries, bond- and vertex-renormalization~\cite{Brout:1960p13109,Horwitz:1961p11045,Englert:1963p11056,Wortis:1974p13094} becomes possible, making computations to very high order possible (currently around 20th order~\cite{Campostrini:2001p12780}.)

In the case of a biological network, however, many of these techniques become impractical; the standard renormalization procedures are frustrated by the strong inhomogeneity in the network, and the unrenormalized graphs are far more numerous and still require computation of the symmetry factors. When a network is characterized by repeating motifs within a larger lattice, however, the enumeration of subgraphs becomes plausible.

In the DCM, to compute the partition function, $Z$, on a graph $G$, we take all subgraphs $g$ of $G$ with vertices even; this set is written $E(G)$ and includes disconnected subgraphs. We can then write
\begin{equation}
Z=2^{N(G)}(\cosh{\beta J})^{n(G)}\sum_{g\in E(G)} v^{n(g)},
\label{zcalc}
\end{equation}
where $n(g)$ is the number of edges in graph (or subgraph) $g$, $N(g)$ the number of vertices, and $v$ is $\tanh{\beta J}$. We take $E(G)$ to include the ``null graph'' with no edges. Finding the derivatives of $Z$ with respect to a set of external fields amounts to allowing some vertices to be odd. We write, for example,
\begin{equation}
P_{a,b}=2^{N(G)}(\cosh{\beta J})^{n(G)}\sum_{g\in E(G,a,b)} v^{n(g)},
\label{pcalc}
\end{equation}
where $E(G,a,b)$ are the subgraphs with all vertices even, when the effective number of edges coming in to vertices $a$ and $b$ are both incremented by one (note that $E(G,a,a)$ is the same as $E(G)$. Then,
\begin{equation}
\langle\sigma_a\sigma_b\rangle=\frac{1}{Z}\frac{\partial^2 Z}{\partial h_i\partial h_j}=\frac{P_{a,b}}{Z},
\end{equation}
and higher-order (connected) correlations yet can be computed as
\begin{equation}
\langle\sigma_{i_1}\sigma_{i_2}\cdots\sigma_{i_k}\rangle = \frac{\partial^k\ln{Z}}{\partial h_{i_1}\cdots\partial h_{i_k}}.
\end{equation}

Direct enumeration of all possible disconnected subgraphs rapidly becomes prohibitive, since computation time is exponential in the number of edges. For the motifs, however, with small $n(M)$ (less than $10$, \emph{e.g.}), the computation can be done on a modern desktop machine. Our general method will be to compose the partition function for $S(q,M)$ from the partition function for $S(q-1,M)$.

\subsection{Branching Networks in the Ising Model} 

Determining the partition function for the branching assembly rule is reasonably straightforward. It is aided by the tree-like hierarchy that arises as the graph is built up; all disconnected, even graphs at any stage can be decomposed into the union of the set of disconnected graphs on the $N(M)$ subgraphs $S(q-1,M)$ and the disconnected graphs on the additional motif $M$ that now forms the ``highest-level'' of the network.
\begin{equation}
Z_q = 2^{N(M)} (\cosh{\beta J})^{n(M)}Z_{q-1}^{N(M)}\sum_{m\in E(M)} v^{n(m)}.
\label{zrp}
\end{equation}
The free energy per vertex, $\ln{Z_q}/N^q$, is a slowly decreasing function of $q$. 

Computing the correlation function of such a system is again aided by the tree-like hierarchy. At stage $q$, copies of the $S(q-1,M)$ graph are placed at the $N(M)$ locations. A vertex $A$ on one of those copies can then be referenced by a string of $q$ numbers $\{a_1,\ldots,a_q\}$, where $a_q$ is the vertex number of $M$ into which the $S(q-1,M)$ graph containing $A$ is placed.

Consider, to begin with, the correlation function between vertex $A$, $\{a_1,a_2\}$, and $B$, $\{b_1, b_2\}$ in $S(2,M)$. When the roots are found on different subgraphs (\emph{i.e.}, $a_2\neq b_2$),
\begin{equation}
\langle \sigma_A \sigma_B \rangle = \frac{1}{Z_2}\frac{\partial^2 Z_2}{\partial h_A\partial h_B} = \frac{1}{Z_2} P_{1,a_1a_2} P_{1,a_2b_2} P_{1,b_2b_1} Z_1^{N(m)-2},
\end{equation}
where $P_{1,ab}$ is the sum of all graphs on $M$ even in all vertices except at vertices $a$ and $b$ which are odd (``subgraphs of $M$ rooted at $a$ and $b$''):
\begin{equation}
P_{1,ab}=\sum_{m\in E(M,\{a,b\})} v^{n(m)}.
\end{equation}
In words, the path from $A$ to $B$ requires leaving the subgraph containing $A$ at $a_2$, crossing $M$, and entering the subgraph containing $B$ at $b_2$. The additional factors of $Z_1$, the partition function on $M$, come from the other subgraphs that, if they are are entered, must be left from the same vertex. The generalization to $n$ roots is straightforward.

The general form for $P$ can be written
\begin{eqnarray}
P_{q,\{a\}\{b\}} & = & P_{q-1,\{a_1...a_{q-1}\}\{a_q...a_q\}} P_{1,a_{q}b_{q}} \nonumber \\
& & \times P_{q-1,\{b_q...b_{q}\}\{b_1...b_{q-1}\}} Z_{q-1}^{N(M)-2}, \label{prp}
\end{eqnarray}
or, in words, that one must get to the most connected node on one's subgraph, and from there travel over the highest-level $M$ to the most connected node of the destination subgraph.

We consider two vertices $\{a\}$ and $\{b\}$ to be separated by a copy distance $d$ where $d$ is the number of subgraphs one must traverse to reach $B$ from $A$ (formally, if $a_d\neq b_d$ but either $d$ is the generation of the graph or $a_{d+1}=b_{d+1}$.) The correlation function has the form of an exponential cutoff:
\begin{equation}
\chi(d)=\frac{1}{|\mathcal{P}(d)|}\sum_{\{A,B\} \in \mathcal{P}(d)} \langle \sigma_A \sigma_B \rangle \sim \chi_0^{d},
\end{equation}
where $\mathcal{P}(d)$ is the set of all vertex pairs separated by copy distance $d$, and $\chi_0$ is the average correlation between different pairs in $M$.

In nesting, local interactions are increasingly less aware of the larger structures in which they are embedded; as copy distance increases, correlations die exponentially. Furthermore, the correlation between two vertices depends only on their relative positions in the hierarchy; the pair is insensitive to the extent of the rest of the graph.

These effects, are due to the way in which the subgraphs are wired together; all interactions between different subgraphs pass through a single-vertex bottlenecks that restrict the number of paths. In the next section, we shall see how the nested construction opens these bottlenecks -- at the cost of larger graph diameters -- and alters the critical behavior.

\subsection{Nested Networks and the General Form}

The branching computations were reasonably simple because of the absence of redundant paths, or loops, above the motif scale. (Formally, the difference -- the set of unshared edges -- between two paths decomposes into a union of even subgraphs on the motif $M$.) The absence of larger redundant paths has many implications in addition to how it affects the correlation functions; for example, connections between distant nodes may be cut by removal of a single vertex.

The nested rule partition function appears harder to compute because of the existence of loops and redundant paths on all scales. However, a general algorithm for the computation of an arbitrary $P_{q,\{a\},\{b\}}$ may be specified. One decomposes the problem into two parts. One first considers how to traverse the ``coarse-grained'' graph, at the highest level; and then considers how to travel ``within'' each coarse-grained vertex to complete the path. The difference between nested and branching then amounts simply to which particular node address on subgraph $A$ allows you to jump to subgraph $B$.

More formally, $P_{q,\{a\},\{b\}}$ is the sum over on the motif $M$ in the following way:
\begin{enumerate}
\item At level $q$, one has a set of roots, $\{a_1,\ldots,a_q\}$, $\{b_1,\ldots,b_q\}$, ... . Each of these roots corresponds to a root in one of the $S(q-1,M)$ copies. For example, the copy number $a_q$ has a root $\{a_1,\ldots,a_{q-1}\}$.
\item Consider in turn each subgraph $m$ in motif $M$ (where $m$ can be disconnected or connected, odd or even).
\item Each edge of that subgraph gives two additional roots, one associated with each end of the edge. For example, an edge between nodes $a_q$ and $b_q$ leads to two new roots, one for the copy $a_q$, and one for the copy $b_q$.
\item If the graph has been constructed by branching, the additional root for the $a_q$ copy is $\{a_q\ldots a_q\}$ (a list $q-1$ entries long.)
\item If the graph has been constructed by nested iteration, the additional root for the $a_q$ copy is  $\{b_q\ldots b_q\}$ (a list $q-1$ entries long.)
\item Thus, for each $S(q-1,M)$ copy, we have a set of roots, $r_i$.
\item Add together $v^{n(m)}$ and the product of the $N(M)$ $P_{q-1, r_i}$.
\end{enumerate}
Note that ensuring the final path is even is deferred to the bottom level, when $P_{1,r_i}$ is computed.

Eq.~\ref{zcalc} is the basis of the direct configurational method; some examples of this expression for small graphs can be found in Ref.~\cite{Sykes:1974p13112}. While enumeration of graphs much larger than 30 bonds is impossible, using the methods described in the text, it is possible to build up much larger graphs with branching and nested properties of interest. With these equations, and Eqs.~\ref{zrp} and \ref{prp}, an arbitrary hierarchy may be constructed, since there is no restriction on the form of $P_{q-1}$.

We have checked the central formulae explicitly through subgraph enumeration on Fig.~\ref{ecoli}; the results for three iterations we have checked through seventh order in $\beta$, and thus in $v$, by an unrenormalized linked-cluster expansion, using Feynman diagrams in the standard fashion~\cite{Muhlschlegel:1963p11388,Wortis:1974p13094}.


\end{document}